\documentclass[aps,prb,reprint,twocolumn,showpacs,superscriptaddress]{revtex4-1}

\usepackage{amsmath, amssymb, braket, dsfont}
\usepackage[mathscr]{euscript}
\usepackage{graphicx}
\usepackage{subfigure}		
\usepackage{epstopdf, times}
\usepackage[breaklinks]{hyperref}

\newcommand{\xiV}{\xi_{V}}			
\newcommand{\shift}{\mathscr{S}}
\newcommand{\chiMPO}{\chi_\textrm{MPO}}
\newcommand{\NLL}{N_\textrm{LL}}

\usepackage[usenames,dvipsnames]{color}			
\definecolor{purple}{rgb}{0.5,0,0.5}

\begin{document}

\title{Infinite DMRG for multicomponent quantum Hall systems}
\author{Michael P. Zaletel}
\affiliation{Department of Physics, University of California, Berkeley, California 94720, USA}
\author{Roger S. K. Mong}
\affiliation{Walter Burke Institute for Theoretical Physics and Institute for Quantum Information and Matter, California Institute of Technology, Pasadena, California 91125, USA}
\author{Frank Pollmann}
\affiliation{Max-Planck-Institut f\"ur Physik komplexer Systeme, 01187 Dresden, Germany}
\author{Edward H. Rezayi}
\affiliation{California State University Los Angeles, Los Angeles, California 90032, USA}

\begin{abstract}

	While the simplest quantum Hall plateaus, such as the $\nu = 1/3$ state in GaAs, can be conveniently analyzed by assuming only a single active Landau level participates, for many phases the   spin, valley, bilayer, subband, or higher Landau level indices play an important role. 
These ``multicomponent'' problems are difficult to study  using exact diagonalization because each component increases the difficulty exponentially.
An important example is the plateau at $\nu = 5/2$, where scattering into higher Landau levels chooses between the competing non-Abelian Pfaffian and anti-Pfaffian states.
We address the methodological issues required to apply the infinite density matrix renormalization group to quantum Hall systems with multiple components and long-range Coulomb interactions, greatly extending  accessible system sizes.
As an initial application we study the problem of Landau level mixing in the $\nu = 5/2$ state.
Within the approach to Landau level mixing used here, we find that at the Coulomb point the anti-Pfaffian state is preferred over the Pfaffian state over a range of Landau level mixing up to the experimentally relevant values.

\end{abstract}

\pacs{73.43.Cd, 05.30.Pr}
\maketitle
\phantomsection
\addcontentsline{toc}{section}{Main text}

\section{Introduction}

Quantum Hall systems have a plethora of experimentally observed phases which have yet to be definitively identified, 
such as the plateaus at  $\nu = 5/2$\cite{Willett:FQH5/2:1987, Eisenstein2002-5/2,Pan1999-5/2-12/5,Xia12/5}
and $\nu = 12/5$,\cite{Pan1999-5/2-12/5,Xia12/5,Manfra12/5} and theoretically proposed (often quite exotic) phases\cite{MooreRead1991,ReadRezayi99,Nayak-2008,BondersonSlingerland2008} which have yet to be observed. 
Numerical simulations, in particular exact diagonalization (ED),\cite{Yoshioka-1983, Haldane-1983, HaldaneRezayi} have long played an important role as a bridge between our experimental and theoretical understanding.
The microscopic physics of these problems often depends on electrons in multiple components, rather than just a single Landau level (LL); this includes the physics of spin, valley degrees of freedom (as in graphene), multilayer systems, and the effect of mixing between higher Landau levels and  subbands. 
For example, if transitions to higher Landau levels are ignored, at $\nu = 5/2$ the Moore-Read state\cite{MooreRead1991} (the ``Pfaffian'') and 
its particle-hole conjugate (the anti-Pfaffian)\cite{Levin-aPf,Lee-aPf} are degenerate and spontaneously break a particle-hole symmetry. \cite{Levin-aPf}
Particle-hole symmetry is lifted if higher Landau levels are included, so ``Landau level mixing'' should play a decisive role in determining which of these two  phases is realized.
While the non-Abelian statistics of quasi-particles of the Pfaffian and anti-Pfaffian phases are similar, they describe different topological phases of matter. In particular, 
their edge structure is quite distinct, which has important implications for the interpretation of interferometry experiments.

	Exact diagonalization (ED) has been successful for certain multicomponent systems,
\cite{McDonaldHaldane:Bilayer2thirds:1996, MooreHaldane:2thirds, Morf98, Ardonne2001, Bilayer1HalfPlus1Half, Bilayer0804.1286,Feiguin-2009, Pairing0807.1034, RezayiSimon-LLMixing, YHWu:2thirds}
but these systems are uniquely difficult because the addition of each component increases the difficulty of exact diagonalization exponentially.
For these multicomponent systems numerical approaches based on the density matrix renormalization group (DMRG) \cite{White-1992, Shibata-2001} may be at a unique advantage.
In this work we explain how the infinite-DMRG \cite{McCulloch-2008, ZaletelMongPollmann} method can be applied to multicomponent fractional quantum Hall (FQH) systems with arbitrary long-range  interactions.
Several of the techniques discussed here, such as the efficient representation of the Hamiltonian and the treatment of long-range interactions, are applicable to DMRG studies of more general two-dimensional (2D) lattice models.
In our implementation of  infinite DMRG for multicomponent systems, holding the amount of entanglement fixed (as would be expected, for example, if we just allow for small amounts of Landau level mixing), the memory complexity of simulating $N$ components scales  as a polynomial $N^3$, in contrast to the exponential scaling of exact diagonalization.

This methodological improvement allows us to simulate between 1\mbox{--}5 components at system sizes well beyond those obtained in exact diagonalization.
For example, keeping the full Hilbert space of 3 spin-polarized LLs at $\nu = 5/2$, we can simulate an infinitely long cylinder of circumference $20 \ell_B$; a comparable $20 \ell_B \times 20 \ell_B$ torus contains $N_\Phi \sim 64$ flux quanta, while ED including mixing into higher Landau levels can access around $N_\Phi \sim 16\mbox{--}20$.\cite{RezayiSimon-LLMixing}
We hope this improvement will find a variety of future applications in the quantum Hall physics of graphene, bilayers, spin-polarization, wide quantum wells, and Landau level mixing.

	This work is organized as follows.
First, we address the methodological issues required to simulate multicomponent FQH systems with long-ranged interactions.
Second, we benchmark our method against exact diagonalization and earlier DMRG studies at filling fractions $\nu = 1/3, 7/3$.
Finally, we study the system at $\nu = 5/2$ with Coulomb interactions;
when keeping a finite number of Landau levels  in the presence of Landau level mixing, there is clear evidence that the anti-Pfaffian (aPf) state is preferred over the Pfaffian (Pf) state. 
While our approach is non-pertubative in the strength of the Landau level mixing, a  truncation of the Hilbert space is required;  the validity of this  approach can be  assessed using complementary methods, which we leave to a future work.

\section{Methods}


While the application of infinite DMRG (iDMRG) to the FQH was discussed in a previous work,\cite{ZaletelMongPollmann} the complexity of a multicomponent system is considerably greater, so we address some methodological aspects we hope will be of use to others.

The infinite DMRG \cite{White-1992,McCulloch-2008,Schollwock2011,Kjall-2012} algorithm finds the ground state of an infinite 1D chain by minimizing  the energy within the variational space of infinite matrix product states (MPS).\cite{Fannes-1992,OstlundRommer1995,RommerOstlund1997}
The accuracy of DMRG is determined by  the bond dimension `$\chi$' of the MPS, which sets the amount of entanglement that can be captured.
In the limit $\chi \rightarrow \infty$ DMRG becomes exact, but the computational difficulty scales as $\mathcal{O}(\chi^3)$.\cite{Hastings-2007,Schuch-2008,Gottesman-2009}
The quantum Hall problem is naturally mapped to a 1D chain by using the Landau-gauge basis of a cylinder of circumference $L$, as will be discussed.
However, since the bipartite entanglement of the cylinder scales linearly with $L$,  the required  bond dimension $\chi$ scales exponentially with the circumference $L$.
Despite this, DMRG can reach larger sizes than exact diagonalization, where the complexity scales exponentially in the \emph{area} of the system.

	To give some sense of the computational resources required, for the $\nu  = 7/3$ data used in Fig.~\ref{fig:thirds}, which does not include Landau level mixing, the largest $L = 24$  data point at $\chi = 15000$  took 3.5 days and 60 GB of memory on a 16 core node.
For the $\nu = 5/2$ data with three Landau levels shown in Fig~\ref{fig:PfaPf},  the maximum bond dimension used was $\chi = 6300$, and a run takes about 2.5 days and 64 GB memory on a single 16 core node. 
In both cases, we store the DMRG `environments' in RAM, so a significant amount of memory could be saved by writing them to disk.

For the long-ranged Hamiltonians studied here, any given site may be involved in $\sim 10^5$ relevant interactions, so evaluating the expected energy  is non-trivial.
One of the key technical achievements of this work is an efficient approximation of $H$ as a matrix product operator (MPO).\cite{Verstraete2004, Crosswhite2008, Pirvu2010}
Given the desired precision \mbox{$\epsilon_\textrm{MPO} > 0$}, we construct a modest-sized MPO to facilitate our computations.
The (memory) complexity of the DMRG algorithm is of order $\mathcal{O}(\chi^2 \chiMPO)$, where $\chiMPO$ is the size of the MPO.
The computational complexity (per DMRG step) scales as $\mathcal{O}(\chi^2\chiMPO^2) + \mathcal{O}(\chi^3\chiMPO)$.
Empirically, we find the scaling $\chiMPO \propto (\log\epsilon^{-1}_\textrm{MPO}) L$, so the main bottleneck lies in $\chi$.

\subsection{Multi-component systems}
	The Coulomb energy $e^2/\epsilon \ell_B$  ($\ell_B$ is the magnetic length)  sets the energy scale in the quantum Hall problem, so we work in units where  $e^2/\epsilon \ell_B = \hbar = 1$. In these units the Coulomb interaction becomes $V(r) = \frac{\ell_B}{r}$.
The splitting between neighboring Landau levels is set by the cyclotron frequency $\omega$.
When the cyclotron energy far exceeds the Coulomb energy $\kappa = 1 / \omega \ll 1$, it is reasonable to study the physics within a single Landau level.
However, in many cases $\kappa$ is not small, leading to scattering between  multiple Landau levels (Landau level mixing), which requires treating a multicomponent quantum Hall system.



\subsubsection{Representation of the Hilbert space}	
\begin{figure}
\begin{centering}
	\includegraphics[width=60mm]{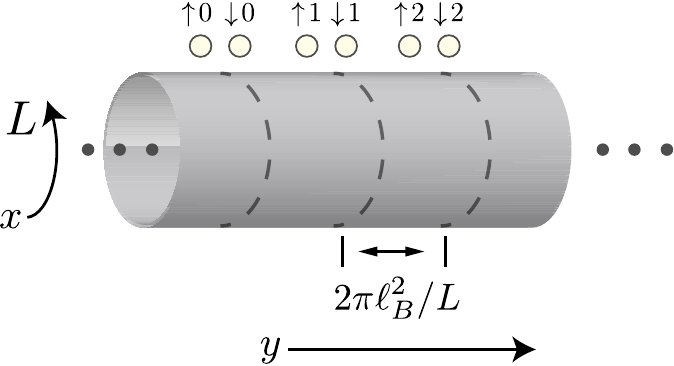}
	\caption{An infinitely long cylinder of circumference $L$.
		The coordinate $x$ runs along the closed direction and $y$ runs along the infinite direction.
		The dotted lines indicate the guiding centers for the orbitals in the Landau gauge, which we map to a 1D chain for DMRG (shown as yellow circles above the cylinder).
		In an $N$-component system, we have $N$ orbitals per guiding center, and thus $N$ sites in the 1D chain per dotted line.
		(Here $N=2$ is illustrated, with  components $\mu = \uparrow, \downarrow$.)}
	\label{fig:cylinder}
\end{centering}
\end{figure}

	We exclusively use the infinite cylinder geometry (Fig.~\ref{fig:cylinder}) as its entanglement properties are best suited to DMRG.\cite{Hu-2012}
The coordinate $x$ runs along the periodic  direction of circumference $L$ while $y$ runs along its infinite length.
The `component' degree of freedom can come from any combination of Landau level index, spin, and valley degrees of freedom, which we label collectively by an index $\mu$.
Working in the Landau gauge $\mathbf{A} = \ell_B^{-2}(-y, 0)$, each component $\mu$ has orbitals labeled by an integer $m$, with momenta $k_x = \frac{2 \pi m}{L}$. 
We work in the full Hilbert space of the multicomponent system; there is no restriction on the occupation within each component.
Each single-particle orbital is labeled by its component and momentum, $\mu m$.
	
	The infinite DMRG algorithm requires an ordering of the single particle states into an infinite 1D fermion chain.
We choose to interleave the components by choosing an order for the $N$ components $\mu_1 < \mu_2 < \cdots < 
\mu_N$ and order the states $\mu \, m$ according to $\mu_1 0 , \mu_2 0, \dots, \mu_N 0, \mu_1 1, \mu_2 1, \dots$. 	
The memory cost of DMRG is linear in the length of the unit cell, so  (holding the amount of entanglement fixed) the multicomponent case leads only to a polynomial increase in complexity.
	
	Due to translation invariance, the most general 2-body interaction is
\begin{align}
	H = \!\!\sum_{r, m, k, \mu, \nu, \rho, \sigma}\!\!  V^{\mu \nu \rho \sigma}_{mk}
		\, \psi^{(\mu)}_{r + 2 m + k} \, \psi^{(\nu)\dag}_{r + m + k} \, \psi^{(\rho)\dag}_{r + m} \, \psi^{(\sigma)}_{r} 
\end{align}
where $V^{\mu \nu \rho \sigma}_{mk}$ are the matrix elements of the 2-body interaction.
At circumference $L$ and interaction range $\xiV$, $V_{mk}$ has contributions out to  $m \sim L$ and $k \sim \xiV L$,
generating $\mathcal{O}(N^4 \xiV L^2)$ non-negligible terms, which amounts to about $10^5$ for the systems studied here.
A compression method is essential.

\subsubsection{Compression of the MPO}	
\label{sec:MPOcompress}

To efficiently store $V$ for the purposes of DMRG we make use of the MPO representation of the Hamiltonian.
The complexity of the DMRG algorithm scales linearly with the size of the MPO $\chiMPO$.

For simplicity we first analyze only the  $m = 0$ terms and drop all component indices.
The interaction takes the form $\sum_{k > 0, r} V_k \hat{X}_{r+k} \hat{Y}_r$, where $\hat{X}$ and $\hat{Y}$ are some one-site operators.
For a set of coefficients $V_k$, the size of the MPO $\chiMPO$ required to \emph{exactly} represent the interaction is generically equal to the number of non-zero values of $V_k$.
One exception to this rule is that a $\chiMPO = 3$ MPO can faithfully represent an exponentially decaying interaction, i.e., $V_k = A^k$ independent of the scalar $|A| < 1$.
The key idea is that we can dramatically decrease $\chiMPO$ by \emph{approximating} the sequence $V_k$ with a sum of exponentials, $V_k \approx \sum_{a=1}^\Lambda B_a (A_a)^k$.\cite{Crosswhite2008, Pirvu2010}
The size of the MPO is then the number of exponentials $\chiMPO = 2 + \Lambda$, which can be far less than the range of the interaction.

This leads to a variational problem in $(A_a, B_a)$ to minimize the error $\big|\!\big| V_k - \sum_{a=1}^\Lambda B_a A_a^k \big|\!\big| = \epsilon_\textrm{MPO}$.
	\footnote{We define the error $\epsilon_\textrm{MPO}$ to be the ``Hankel norm'' (i.e., largest Hankel singular value) for the difference $\delta_k = V_k - \sum_{a=1}^\Lambda B_a A_a^k$; with this definition the technique we use is truly optimum.\cite{Pirvu2010}
		Qualitatively $\epsilon_\textrm{MPO}$ behaves like a regular norm.}
Naturally $\epsilon_\textrm{MPO}$ decreases with larger $\Lambda$; for the quantum Hall potentials we observe a modest scaling $\Lambda \sim \mathcal{O}({-\log\epsilon_\textrm{MPO}})$.

Once including multiple components and the range of $m$, the quantum Hall problem is even more complex.
Each integer $m$ can be analyzed in isolation, and requires a decomposition of the form
\begin{align}
	V_{mk}^{\mu \nu \rho \sigma} \approx \begin{cases}
			\displaystyle\sum^{\Lambda_m}_{a, b=1} C^{\mu \nu ; a}_{m} \big(A_m^{k-1}\big)_{a;b} B^{b ; \rho \sigma}_{m} & k > 0 ,
		\\	\qquad\qquad D_m^{\mu \nu ; \rho \sigma} & k = 0 .
		\end{cases}
	\label{eq:Vdecomp}
\end{align}
For each fixed $m$, $A_{m}$ is a $\Lambda_m \times \Lambda_m$ matrix with indices $(a; b)$; $C_m$ is a $N^2 \times \Lambda_m$ dimensional matrix with indices $(\mu \nu; a)$, and $B_m$ is $\Lambda_m \times N^2$ dimensional matrix with entries $(b; \rho \sigma)$. 
$A_m^k$ denotes the $k$\textsuperscript{th} power of $A_m$.
Given the matrices $(A, B, C, D)_m$ it is trivial to construct an MPO representation for Eq.~\eqref{eq:Vdecomp}.
For a concrete example of how $(A, B, C, D)$ can be used to construct the MPO, we refer to App.~\ref{app:mpo_example}.
The size of the MPO (and hence the numerical difficulty) scales as $\chiMPO \approx \sum_m (\Lambda_m + 2)$.
The interactions decay as $V_{mk} \sim e^{-(2 \pi m / L)^2 }$ at large $m$, so only $|m| \sim \mathcal{O}(L)$ sectors need to be kept.

MPO compression of this form leads to tremendous gains in efficiency.
For an $N$-component system at circumference $L$, the dimension $\chiMPO$ of the uncompressed MPO (as used in our last study Ref.~\onlinecite{ZaletelMongPollmann}) scales as $\chiMPO \sim \mathcal{O}\big(N^4 L^2 / (2\pi\ell_B)^2\big)$, which becomes prohibitively expensive.
In contrast, the compressed MPO scales as $\chiMPO \sim \mathcal{O}\big(N^2 L/(2\pi\ell_B)\big)$.
For example,  a Coulomb interaction between 3LLs requires an unoptimized MPO of dimension $\chiMPO \approx 5000$ at $L = 20 \ell_B$,  but only $\chiMPO \approx 400$ with compression.

For each $m$, we wish to find the matrices $(A, B, C, D)_m$ which best approximate $V$ given the finite rank $\Lambda_m$.
Luckily finding optimal approximations of this form is a well studied problem in control theory called ``model reduction.''
Fixing $m$, we can view $V_{mk}^{\mu \nu \, \rho \sigma} $ as the signal of a multiple input, multiple output  discrete state space machine, where $\mu \nu$ label the $N^2$ `inputs,' $ \rho \sigma$ label the $N^2$ `outputs,' and $k = 1, 2, \dots$ plays the role of `time.'\cite{wikiSSR}
The signal $V_k$ is viewed as the Green's function  (alias the ``transfer function'') of a linear process whose dynamics are governed by Eq.~\eqref{eq:Vdecomp}.
We wish to best approximate this $N^2$-input, $N^2$-output signal with a rank $\Lambda_m$ state space state machine. 
Our notation $A, B, C, D$ reflects the standard control theory notation.

The optimal $(A, B, C, D)_m$  can be found using a technique called the  block-Hankel singular value decomposition.\cite{KungLin1981} 
In the control systems literature, the resulting state space machine is encapsulated in a block matrix $\left(\!\begin{smallmatrix}A & C \\B & D\end{smallmatrix}\!\right)_{\!m}$, which is the desired  data of Eq.~\eqref{eq:Vdecomp}.
While the Hankel method is straightforward to implement in the single component (1-input, 1-output) case, the open source SLICOT library can conveniently turn the signal $V$ into the optimal representation $(A, B, C, D)$ in the general case.\cite{slicot}

\subsubsection{Validation with model Hamiltonians and exact diagonalization}
\begin{figure}
\begin{centering}
	\includegraphics[width=70mm]{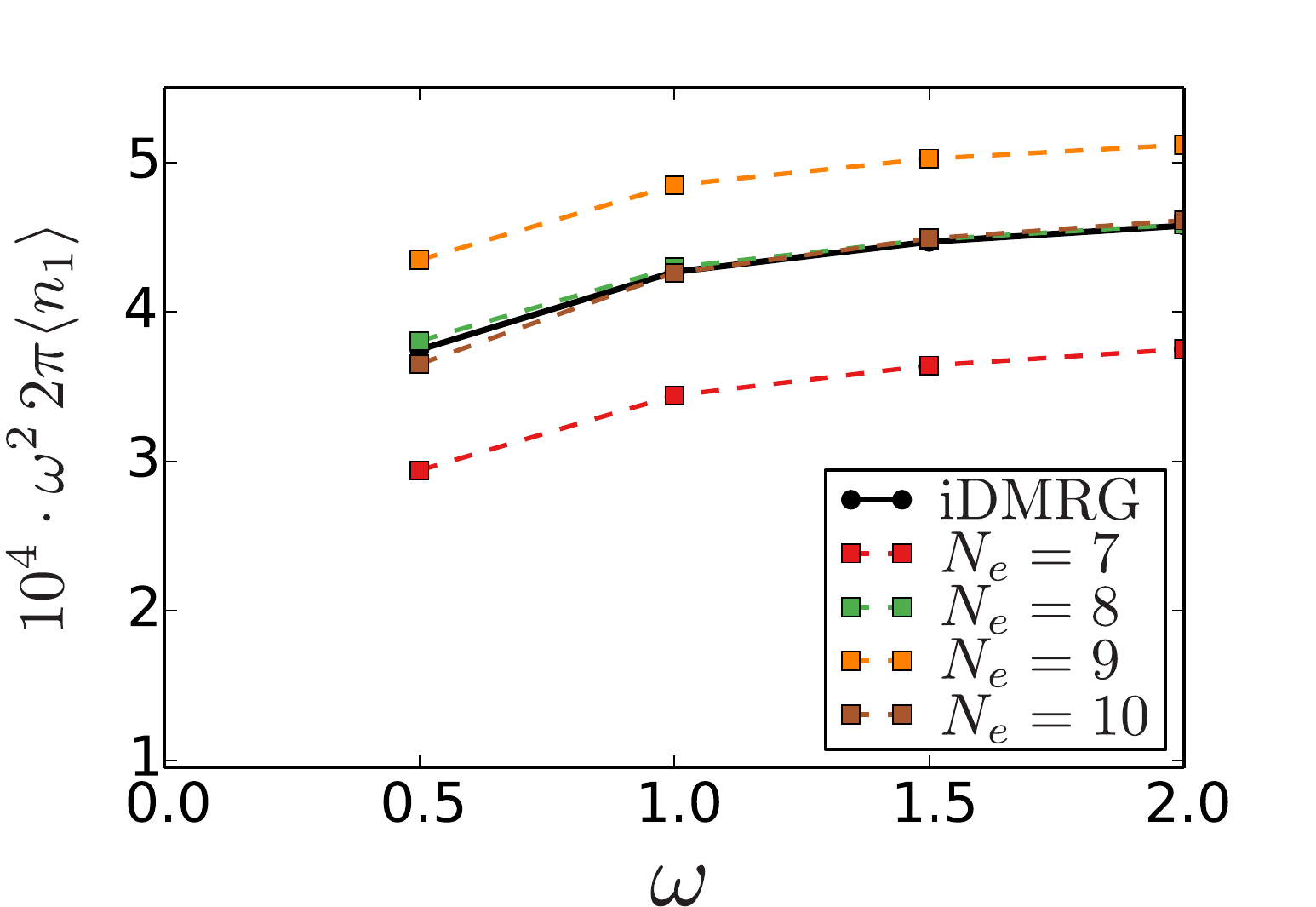}
	\caption{%
		Occupation of the $n = 1$ Landau level as a function of the Landau level splitting $\omega$, in units with $\hbar = e^2 / \epsilon \ell_B = 1$.
		Lowest order perturbation theory in $\omega^{-1}$ predicts $\braket{\hat{N}_1} \sim \omega^{-2}$, which we verify for $\omega \gg 2$, but modest higher order effects appear near $\omega \sim 1$, the regime of physical interest.
		iDMRG at $L = 17\ell_B$ (solid) is in good agreement with exact diagonalization (dashed lines).
	}
	\label{fig:ED}
\end{centering}
\end{figure}

	Due to the complexity of implementing a multicomponent Hamiltonian, we have checked our implementation using a both model interactions and exact diagonalization.
First, we consider filling $\nu = 2/5$ with  hard-core interaction $V(q) = -q^2$. The Jain state is an exact zero-energy eigenstate when Landau level mixing is allowed between the $n = 0, 1$ levels at vanishing cyclotron splitting $\omega = 0$. We have verified that iDMRG finds a zero-energy state to arbitrary precision as the accuracy of the MPO compression is increased.

	Second, we consider filling $\nu = 1/3$ with Landau level mixing between levels $n = 0, 1$.
To minimize finite size effects, we use a finite range potential $V(r) = e^{- r^2 / 8\ell_B^2} \ell_B/r$.
We compare iDMRG on a cylinder of circumference $L = 17\ell_B$ with exact diagonalization of $N_e = 7, 8, 9, 10$ electrons on a square  torus. 
As an observable we measure the average occupation of the $n = 1$ Landau level $\braket{\hat{N}_1}$ as the cyclotron energy $\omega$ is increased.
Results are reported in Fig.~\ref{fig:ED}, which oscillate about the iDMRG values  and are clearly consistent, and in good agreement with them. 

\subsection{Long range interactions}
	Previous work\cite{ZaletelMongPollmann} on the infinite cylinder was limited to short-range interactions, which we now address using the optimized MPOs and an extrapolation procedure.
On a finite-size system various definitions of the Coulomb interaction are possible, such as replacing distance by a chord length.
In this work we regulate the interaction with a Gaussian envelope,
\begin{align}
	V(r) = \frac{\ell_B}{r} e^{-\tfrac{1}{2} (r/ \xiV)^2}
	\label{eq:Coulomb_Gaussian}
\end{align}	
and periodize the interaction in the compact direction $x$.
While the Coulomb tail will have important effects when the density is not uniform, such as for striped phases or excitations, for a gapped uniform ground state the energetics beyond a correlation length $r > \xi$ are purely direct, i.e.,  the exchange energy is negligible.
Hence for large $r$ the energy is simply $E \sim V(r) \langle\rho(x) \rangle \langle \rho(x+r) \rangle$, which is accounted for by subtraction of the $V(q = 0)$ component of the energy.

	However, for $r < \xi$ correlations are important, and by Taylor expanding the Gaussian envelope we expect an effect at short distances of order $\mathcal{O}(1/\xiV^2)$. This motivates the following extrapolation for the ground state energy:
\begin{align}
	E(\xiV) = E_0 + a \xiV^{-2} + \cdots
	\label{eq:int_extrapolation}
\end{align}
which is valid so long as $\frac{L}{2} > \xiV > \xi$.
Once the energy can be fitted to this form we can assume the long range interactions are purely direct and reliably extrapolate to the Coulomb interaction.

\begin{figure}
	\includegraphics[width=79mm]{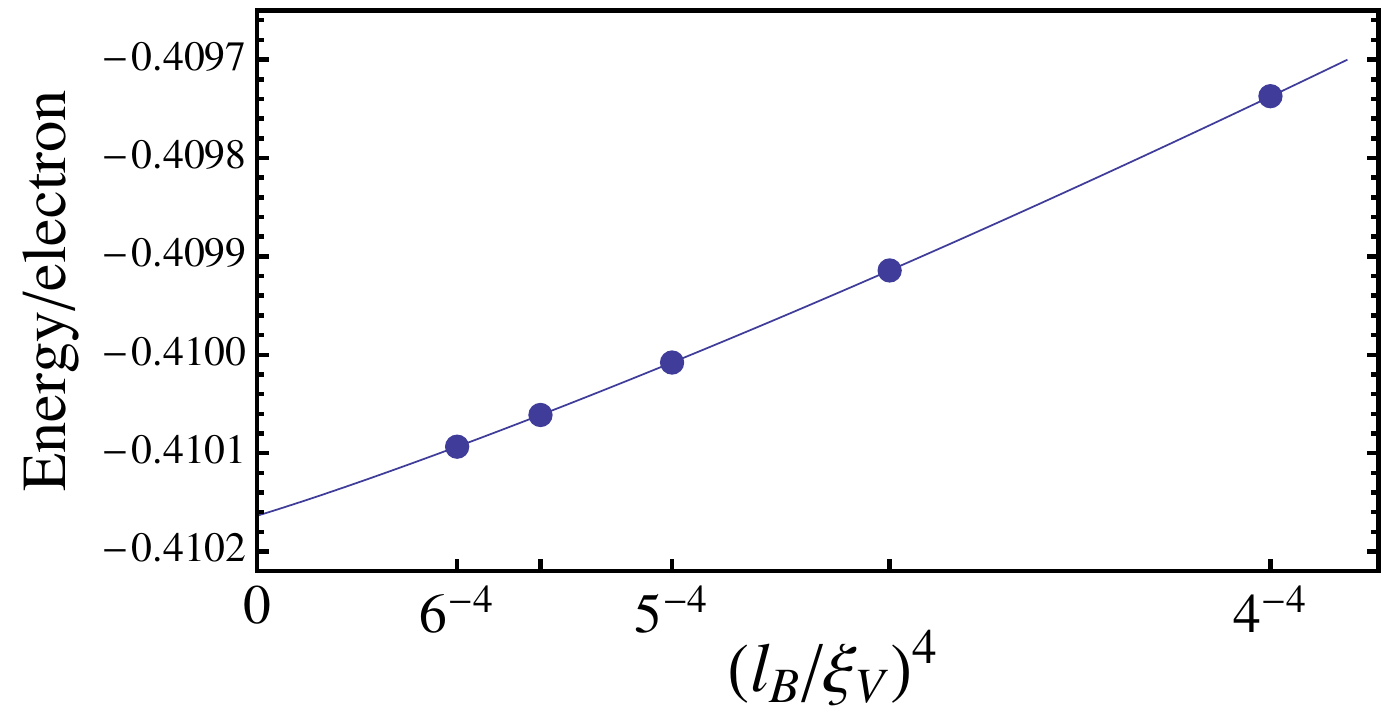}
	\caption{%
		Extrapolation of the energy per electron at $\nu = 1/3$ with interaction given by Eq.~\eqref{eq:Coulomb_xiorder1} for $n=1$.
		The energy $E(\xiV)$ is plotted as a function of $\xiV^{-4}\ell_B^4$, where $\xiV$ is the ``cutoff'' length scale.
		Extrapolating the energy to $\xiV \rightarrow \infty$ via the functional form Eq.~\eqref{eq:xiorder1_E}, we find $E(\infty) = -0.410164(4)$ for the Coulomb potential, in good agreement with previous DMRG study of $E = -0.41016(2)$.\cite{Zhao-2011}
	}
	\label{fig:E_one_third}
\end{figure}

To validate this procedure we have calculated the ground state energy at the $\nu = 1/3$ Coulomb point, where accurate energies can also be obtained using exact diagonalization and  finite-sphere DMRG.\cite{Zhao-2011}
For the purpose of very accurately computing the ground state energy, we use a higher order generalization of the Gaussian envelope,
\begin{align}\begin{split}
	V(r) &= \frac{\ell_B}{r} f_n\Big(\frac{r}{\xiV}\Big) ,
\\	f_n(z) &= e^{-z^2/2} \sum_{m=0}^n \frac{1}{m!} \bigg( \frac{z^2}{2} \bigg)^m .
	\label{eq:Coulomb_xiorder1}
\end{split}\end{align}
The enveloping function $f_n$ is constructed to give higher order approximations to unity at small distance, $f_n(z) = 1 - \mathcal{O}(z^{2n+2})$, and thus a better extrapolation to the ground state energy.
For the first order envelope $f_1$, the energy takes the form
\begin{align}
	E(\xiV) &= E_0 + a_2\xiV^{-4} + a_3\xiV^{-6} + \dots	&& (n=1)
	\label{eq:xiorder1_E}
\end{align}
Figure~\ref{fig:E_one_third} shows the ground state energy obtained via DMRG, and its fit to the functional form above.
Each data point $E(\xiV)$ is obtained via an extrapolation to the $L \rightarrow \infty$ limit.
We obtain $E_0 = -0.410164(4)/\textrm{electron}$ (in units of $\frac{e^2}{\epsilon \ell_B}$), in excellent agreement with previous finite DMRG results of $E = -0.41016(2) $. \cite{Zhao-2011}

In the remainder of this paper, we will use only the Gaussian envelope given by Eq.~\eqref{eq:Coulomb_Gaussian}.

\subsection{Ergodicity of the iDMRG algorithm}
DMRG is a local optimization algorithm, so for longer-range Hamiltonians it is susceptible to getting stuck.
In fact, one can prove the standard two-site DMRG algorithm cannot explore the full variational space, so it will not find an optimal MPS.
Reference~\onlinecite{ZaletelMongPollmann} overcame this difficulty by using an $n$-site algorithm for some $n > 2$ that depended on the filling.
However, this significantly increases the memory requirements of the algorithm.
In an earlier finite DMRG study, \cite{Feiguin-2009} it was shown that White's `density matrix corrections'  can also overcome this problem. \cite{White2005}
The density matrix corrections can also be implemented in infinite DMRG, and we find that the two-site algorithm combined with density matrix corrections does avoid sticking, so 
is used exclusively here.

\subsection{Characterization of quantum Hall phases}
In addition to energetics and local observables (such as structure factors), iDMRG is well suited to determine the topological order of a state.
Here we briefly review the set of measures used in this work.

	The most basic fingerprint of topological order is a protected ground-state degeneracy on a cylinder, with one degenerate state per anyon in the theory.\cite{Wen:TopOrder:1990}
There is a special basis---the minimally entangled basis\cite{YZhangMES2012}---in which each basis state is in correspondence with an anyon `$a$' in the topological field theory, so we label the ground states as $\ket{\Psi_a}$ by anyons types.
The entanglement properties of each ground state $a$ reveal a remarkable amount of universal information about the anyons.
The starting point of these measures is the Schmidt decomposition.
Splitting the Hilbert space into the orbitals to the left and right of some bond,
	\footnote{In fractional quantum Hall, translation is equivalent to threading anyon flux through the cylinder, and thus the entanglement measures will depend on the choice of bond dividing the system.
		We fix a particular bond (between sites 0 and 1) as a convention.}
	$\mathcal{H} = \mathcal{H}_L \otimes \mathcal{H}_R$, a state can be decomposed as	
\begin{equation}
	\ket{\Psi_a} = \sum_{\alpha} \lambda^{(a)}_{\alpha} \ket{\alpha^{(a)}}_L \ket{\alpha^{(a)}}_R.
\end{equation}
For each wavefunction $\ket{\Psi_a}$, the Schmidt states $\ket{\alpha^{(a)}}_{L/R}$ form orthonormal bases for the sites to the left/right of the cut, and $\lambda^{(a)}_{\alpha}$ are the Schmidt values.
The ``entanglement spectrum'' is the collection of Schmidt values $ \lambda^{(a)}_{\alpha}$, which can be trivially calculated from the MPS used in the DMRG method.
The entanglement entropy for this bipartition is directly obtained from the Schmidt values $S_{a} = -\sum_{\alpha} (\lambda^{(a)}_{\alpha})^2 \log (\lambda^{(a)}_{\alpha})^2$.

	The first entanglement measure is the topological entanglement entropy $\gamma_a$.\cite{KitaevPreskill,Levin-2006}
The entanglement entropy of a ground state $a$ should scale with the circumference $L$ as
\begin{align}
	S_a(L) &= \beta L - \gamma_a + \dots, \qquad \gamma_a = \log(\mathcal{D}/d_a),
	\label{eq:TEE}
\end{align}
where $\beta$ is a (non-universal) constant independent of anyon type $a$, and the ellipsis denotes terms decaying with $L$.
$d_a$ is the quantum dimension of anyon $a$, and $\mathcal{D} \equiv \sqrt{\sum_a d_a^2}$ is the total quantum dimension of the theory.

Furthermore, quantum Hall systems on a cylinder have a conserved momentum $K$ corresponding to rotations of the cylinder.
Each orbital $\mu m$ has momentum $2 \pi m / L$, so we define a scaled momentum operator  $\hat{K}$,
\begin{subequations}\label{eq:charges}\begin{align}
	\hat{K} &= \sum_{\mu,m} \hat{K}_{\mu,m} \equiv \sum_{\mu, m} m ( \hat{N}_{\mu,m} - \nu_\mu ) \quad \textrm{(momentum)}	,
\end{align}\end{subequations}
where $\hat{N}_{\mu, m}$ is the number operator at site $\mu, m$, and $\nu_\mu$ is the average filling of component $\mu$. 
For any cut $L/R$, the momentum is a sum of the momenta to the left/right of the cut, $\hat{K} = \hat{K}_L + \hat{K}_R$.
Each left Schmidt state has definite momentum,
\begin{align}
	\hat{K}_L \ket{\alpha^{(a)}}_L = K^{(a)}_{\alpha}  \ket{\alpha^{(a)}}_L. 
\end{align}
Remarkably, the `entanglement average' of the momenta $K^{(a)}_\alpha$s within ground state $a$
\begin{align}
	\braket{K^{(a)}} \equiv \sum_{\alpha} (\lambda^{(a)}_{\alpha})^2 K^{(a)}_{\alpha}
\end{align}
encodes topological information. 
While we refer to Ref.~\onlinecite{ZaletelMongPollmann} for the details, there exists a simple quantity $K_\textrm{orb}(L)$ which can be computed from the filling of each component%
	\footnote{The orbital portion of the momentum polarization is given by $K_\textrm{orb}(L) = -\frac{\nu}{24} - \big(\frac{L}{4\pi\ell_B}\big)^2 \sum_\mu \nu_\mu(2n_\mu+1)$, where $\mu$ runs over the QH components, $\nu_\mu$ are their filling factors which totals to $\nu = \sum_\mu \nu_\mu$, and $n_\mu$ are the Landau level indices.
		This term exists because the orbitals have finite extent along the $y$-direction.}
	such that
\begin{align}\begin{split}
	P_a(L) &\equiv \braket{K^{(a)}} (L) + K_{\textrm{orb}}(L)
	\\	&= -\frac{\shift\,\nu}{(4\pi\ell_B)^2} L^2 + h_a - \frac{c}{24} + \ldots \pmod1,
	\label{eq:mompol}
\end{split}\end{align}
where once again the ellipsis denotes terms decaying with $L$.
Here $\shift$ is the ``shift,'' an integer invariant related to the Hall viscosity;\cite{Tokatly-Vignale-2007,*Tokatly-Vignale-2007Err,Read-2009}
$c$ is the chiral central charge of the edge states;
and $e^{2 \pi i h_a} = \theta_a$ is the topological spin of anyon $a$.
In a subsequent work applying the same concepts to lattice systems, $P_a$ was called the ``momentum polarization.''\cite{HHTu:Mompol}

In addition to the aggregate quantities $S_a$ and $P_a$ obtained from the Schmidt weights, the level structure of the entanglement spectrum itself contains information regarding the excitation spectrum of a physical edge.\cite{KitaevPreskill, LiHaldane, QiKatsuraLudwig}
Plotting the ``entanglement energies,'' defined by $E_\alpha \equiv -\log\lambda_{\alpha}^2$, organized by their momentum eigenvalues $K_\alpha$, generically provide a fingerprint for the topological phase.
This method complements the other approaches described here.
 
In summary, knowing only the entanglement spectrum as a function of the circumference, we can capture a remarkable amount of data: $d_a$, $h_a$, $c$, $\shift$, as well as the edge structure.
For practical purposes this is enough to distinguish between competing topological orders.

Finally, we also calculate the correlation length $\xi$, computed via the transfer matrix of the MPS.\cite{Schollwock2011}
The quantity $\xi$ provides an upper bound to the decay length for \emph{all} ground state correlation functions along the length of the cylinder.
While $\xi$ is not a topological invariant, it carries useful information and may serve as a proxy for the size of quasiparticles.

\begin{figure*}
	\begin{tabular}{cc}
		\subfigure[\quad$\nu=1/3$]{ \includegraphics[width=80mm]{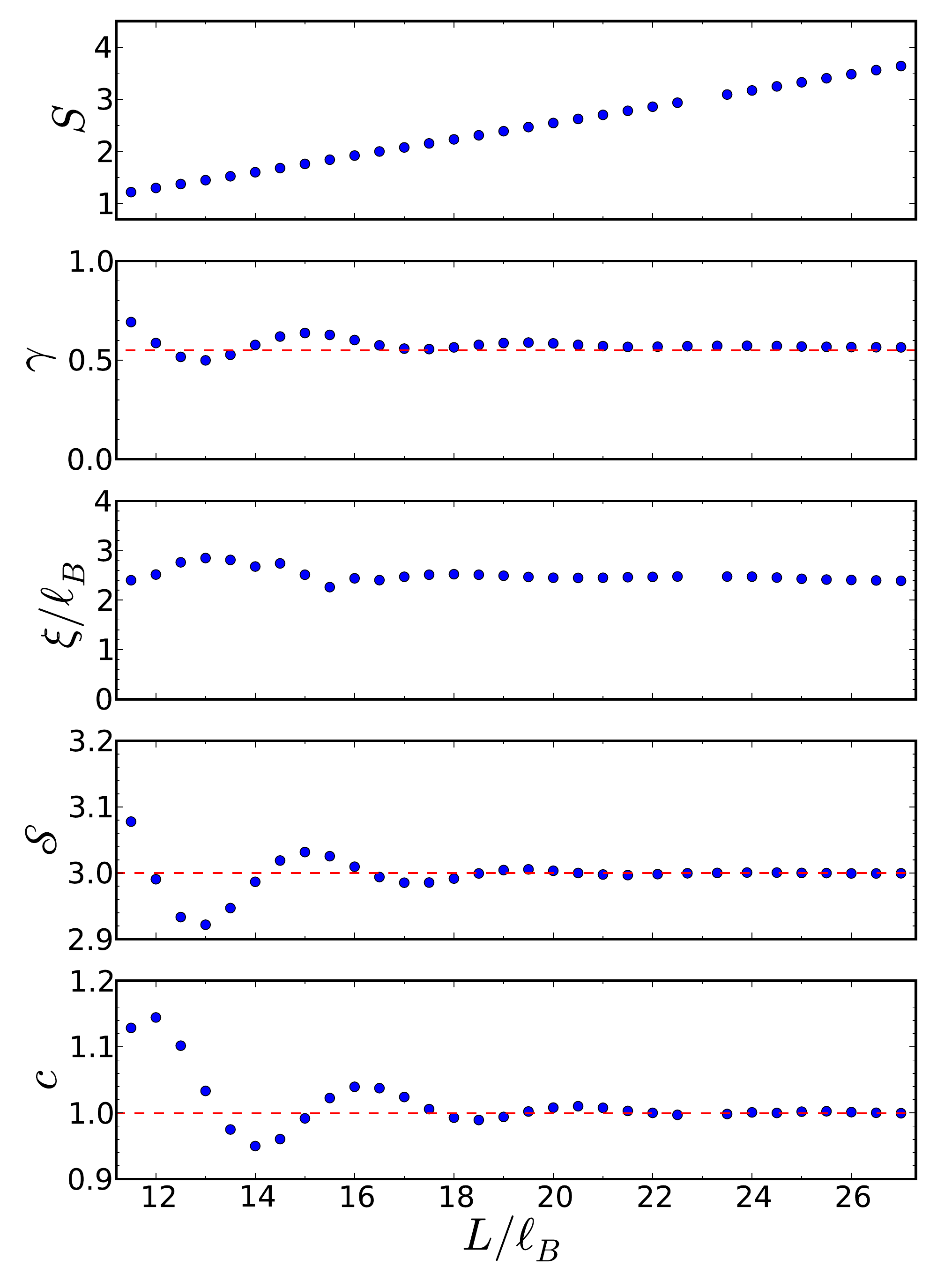}\quad\quad \label{fig:1third} }
	&	\subfigure[\quad$\nu=7/3$]{ \includegraphics[width=80mm]{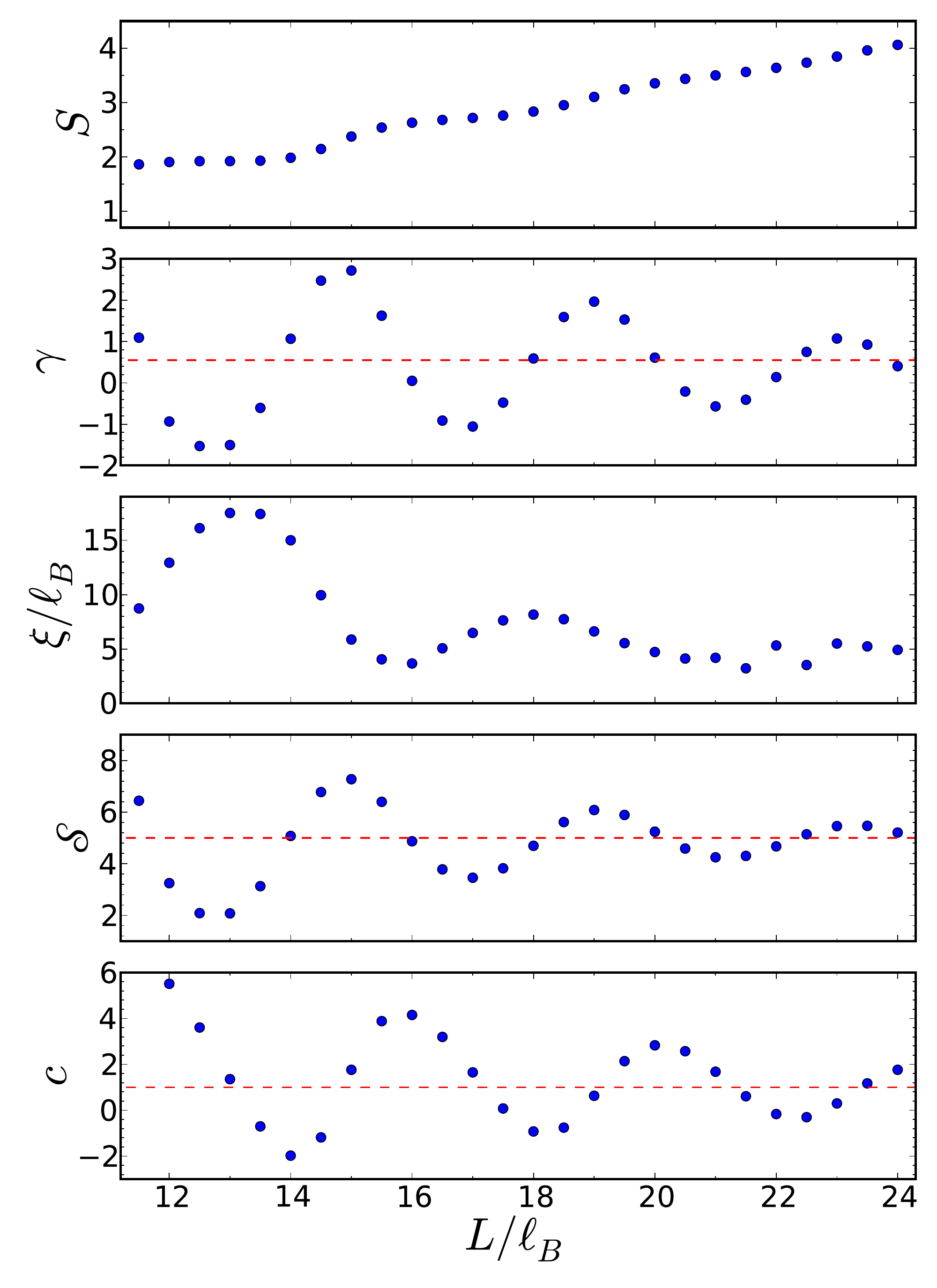}\quad\quad \label{fig:7thirds} }
	\\	\subfigure[\quad$\nu=1/3$]{ \includegraphics[width=75mm]{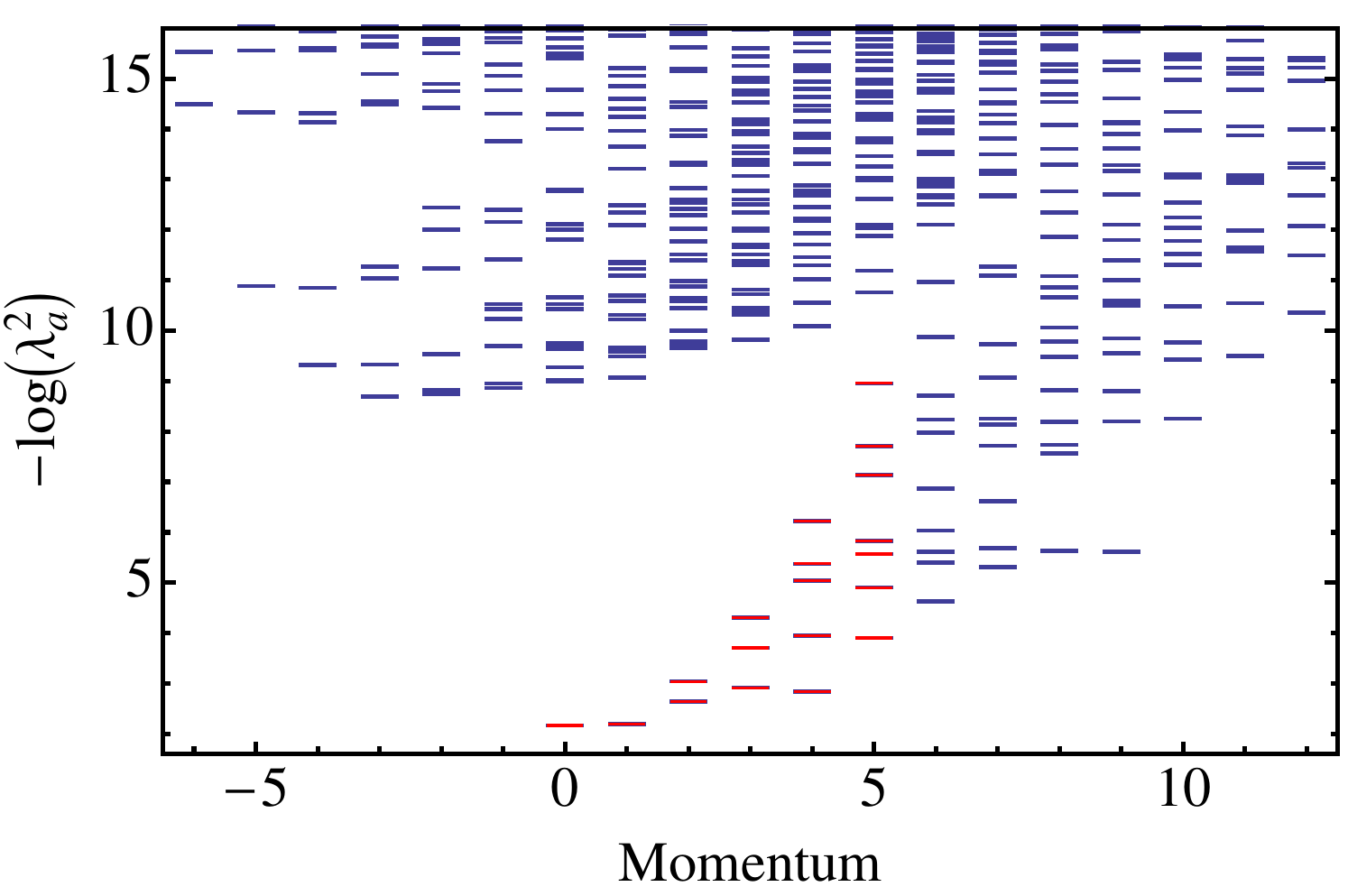}\qquad \label{fig:es1third} }
	&	\subfigure[\quad$\nu=7/3$]{ \includegraphics[width=75mm]{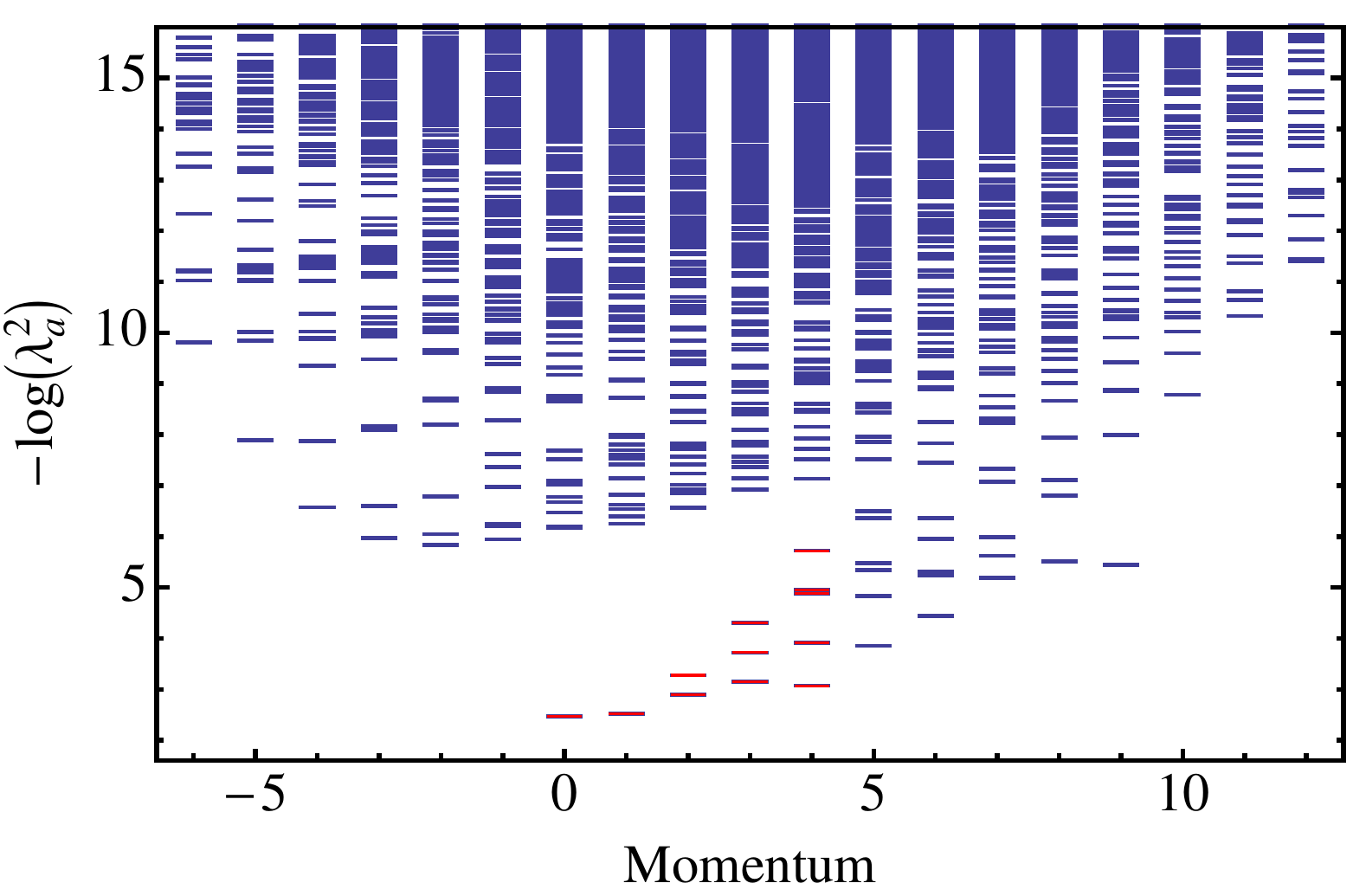}\qquad \label{fig:es7thirds} }
	\end{tabular}
	\caption{%
		(Top) Plot of various quantities as a function of circumference $L$ for the one third-filled spin-polarized state in the zeroth ($\nu=1/3$) and first ($\nu=7/3$) Landau level, with no Landau level mixing.
		The simulations were done at $\xi_V = 5$ with MPS bond dimensions up to $\chi = 15000$.
		The five plots are, from top to bottom, entanglement entropy ($S$), topological entanglement entropy ($\gamma = L\frac{dS}{dL}-S$), correlation length ($\xi$), shift ($\mathscr{S}$), and chiral central charge ($c$), with the theoretical values given by the red dashed lines.
		The oscillations are far more pronounced in the $\nu=7/3$ state than in the $\nu=1/3$ state.
		(Bottom) Orbital entanglement spectra for the two states at $L = 25\ell_B$, organized by their momenta.
		Here only Schmidt states in the neutral charge sector (fixed number of particles on each side of the entanglement cut) are shown.
		The ``low energy'' portion of the spectra is highlighted.
	}
	\label{fig:thirds}
\end{figure*}
	
\section{\texorpdfstring{Comparison of the Coulomb point at fillings $\nu = 1/3$ and $\nu = 7/3$}{1/3}}
	Given the robustness of the $\nu = 1/3$ Coulomb phase and its well behaved entanglement properties, the nature of the $\nu = 2 + 1/3 = 7/3$ Coulomb phase has been remarkably difficult to pin down.\cite{Balram2013, Johri:7thirdsDMRG:2014}
While previous studies generally agree the $\nu = 7/3$ state has the same Laughlin-type order as $\nu = 1/3$, it has been impossible to obtain sharp entanglement measures, such as the topological entanglement entropy.

Taking advantage of our treatment of long-range interactions, we have applied  infinite DMRG to the spin-polarized Coulomb point at $\nu = 7/3$. 
We ignore the effect of Landau level mixing.
In Fig.~\ref{fig:thirds}, we compare various topological measures as a function of cylinder circumference $L$ for $\nu = 1/3, 7/3$.
The topological quantities $\gamma$, $\shift$ and $c$ are extracted via fits to Eqs.~\eqref{eq:TEE} and \eqref{eq:mompol}.
The red dashed line indicates the expected theoretical values for a Laughlin state in the zeroth $\nu=1/3$ and first $\nu=7/3$ Landau levels.
In both cases we expect $\gamma = \log\sqrt{3} \approx 0.55$, $c = 1$, with $\shift = 3,5$ for $\nu=1/3,7/3$ respectively.

The scalings of $S_a(L)$ and $P_a(L)$ have non-universal, exponentially decaying corrections, so a fit must be used to extract the universal components. 
In Figs.~\ref{fig:1third} and \ref{fig:7thirds}, we use windowed fits for $\gamma$ and $\shift$, which provide guidance on the convergence and reliability of the data.
For data points taken at circumferences $\{ L_1, L_2, \dots \}$, we choose a small subset of data points centered at some $L$, and fit the subset to the functional forms of Eqs.~\eqref{eq:TEE}, \eqref{eq:mompol} with no further subleading terms.
This gives an estimate of the desired invariants at system size $L$, and convergence can be checked as a function of circumference.

In addition to the non-universal subleading terms, the leading coefficient  $\alpha$ for the entanglement entropy Eq.~\eqref{eq:TEE} is also non-universal, and thus extracting the topological entanglement entropy $\gamma$ is subject to severe extrapolation errors.
On the other hand, the shift $\shift$ can often be reliably extracted as it constitutes the \emph{leading} term of Eq.~\eqref{eq:mompol}, and is guaranteed to be integer-valued for isotropic phases.
For these reasons, the shift converges rapidly with system size, while it is difficult to get a precise value of $\gamma$.

The correlation length of the $\nu = 1/3$ state is measured to be $\xi = 2.5\ell_B$, while for $\nu = 7/3$, $\xi \sim 5\ell_B$.
The increased length scale at $\nu = 7/3$ is in agreement with previous studies on the size of the quasiparticle excitations.\cite{Balram2013, Johri:7thirdsDMRG:2014}
Despite this rather modest difference in correlation length, when using the windowed fit procedure the amplitude of the oscillatory behavior at $\nu = 7/3$ is 10\mbox{--}50 times more severe than that of $\nu = 1/3$.
While the results are all consistent with Laughlin-type order, it is not possible to accurately extract the entanglement measures even at a circumference $L = 25\ell_B$, which is five times the correlation length $\xi \approx 5\ell_B$.
The period of oscillations in both states is $\ell/\ell_B \sim 4.2\mbox{--}4.6 $.
It may be that the wavefunction of the $\nu = 7/3$ states has a higher amplitude for Wigner-crystal like configurations, which are frustrated at incommensurate $L$, leading to the observed oscillations. 

	This finding illustrates that the physical correlation length is not a reliable guide to the convergence of entanglement properties. 
There is no rigorous reason why the length scales in the entanglement spectrum that governs the exponential converge of topologically protected properties should be those of the physical system.
Indeed, perverse examples can be constructed\cite{BravyiRumour} for which the entanglement length scale diverges even while the physical correlation length is unchanged.
This is worth keeping in mind for a variety of DMRG studies  which require finite-circumference extrapolation.

Despite the poor convergence of the topological entanglement invariants $\gamma$, $c$, $\shift$ for the $\nu = 7/3$ state,
the entanglement spectrum provides very strong evidence in favor of a Laughlin phase.
In Figs.~\ref{fig:es1third} and \ref{fig:es7thirds}, we plot the entanglement energies organized by their momentum eigenvalues $K_\alpha$.
(The data presented are taken for the identity $a = \mathds{1}$ sector.)
The ``low energy'' portion (large $\lambda_\alpha$'s) shows the characteristic Laughlin-state counting 1, 1, 2, 3, 5, etc.\ at both filling fractions.


\section{Analysis of \texorpdfstring{$\nu = 5/2$}{5/2} with Landau level mixing}
	The plateau observed at filling fraction $\nu = 5/2$ is a potential host of non-Abelien anyons,
raising the stakes in the search for an experimental signature of non-Abelian statistics.\cite{MooreRead1991,Nayak-2008,BondersonNonAbelianProof2011}
Edge interferometry experiments could in principle detect non-Abelian statistics.\cite{DasSarma,BondersonInterferometer-2006,Nayak-2008,BisharaInterferometry-2009}
In practice the edge could be messy, making the interpretation difficult. These issues are beyond the scope of this paper, which will be addressing the bulk physics.
Temporarily ignoring the effects of Landau level mixing, $\nu = 2$ of the electrons fill the lowest Landau level with both spin species, while the remaining $\nu = 1/2$ reside in the $n=1$ Landau level.
If the  $\nu = 1/2$ component is spin-polarized, the most likely candidate phases are the non-Abelian Pfaffian and anti-Pfaffian states.
If the $\nu = 1/2$ is unpolarized, a Halperin 331 bilayer state is likely.\cite{Halperin:FQHHierarchy:1983, ReadRezayi:PairedQH:1996}

Experimental evidence for spin-polarization has been mixed at best.
Most recent experiments, employing different methods, find 
polarized states,\cite{Tiemann,Stern-2012} unpolarized states,\cite{Stern-2010,Pinczuk} or, depending on the density, both.\cite{Pan}
Numerically, previous exact diagonalization studies give strong evidence for spin-polarization;\cite{Morf98,Feiguin-2009,RezayiSimon-LLMixing} accordingly, in this work we proceed under the assumption of spin polarization.

	While it is often justified to treat the filled Landau levels as inert and drop terms in the interaction which scatter electrons into higher Landau levels, Landau level mixing is crucial at $\nu = 5/2$.
This is because when LL mixing is ignored, a 2-body potential has a particle-hole (PH) symmetry which exchanges filled and empty orbitals in the $n=1$ LL.
The Pfaffian (Pf) state breaks PH  symmetry, and is in fact topologically distinct from its PH conjugate, the anti-Pfaffian (aPf) state.
In the absence of LL mixing the ground state must spontaneously break PH symmetry if it forms either the Pf or aPf.
However, the ratio of the Coulomb energy to cyclotron frequency $\omega$ is of order 1,  $\kappa \equiv e^2 / \epsilon \ell_B \omega \sim 0.7\mbox{--}1.8$, and the resulting scattering into other LLs  breaks PH symmetry.
While LL mixing is in some sense `small' (we find less then 1\% occupation in the higher LLs), the effect is important as it acts as a PH breaking field which differentiates between the Pf and aPf.
The question is what the sign is. 
Experimentally, the presence (absence)  of a neutral counter-propagating edge 
mode would only rule out the Pf (aPf).

\begin{figure*}[t]
	\subfigure[\quad 2 Landau levels]{ \includegraphics[width=7.2cm]{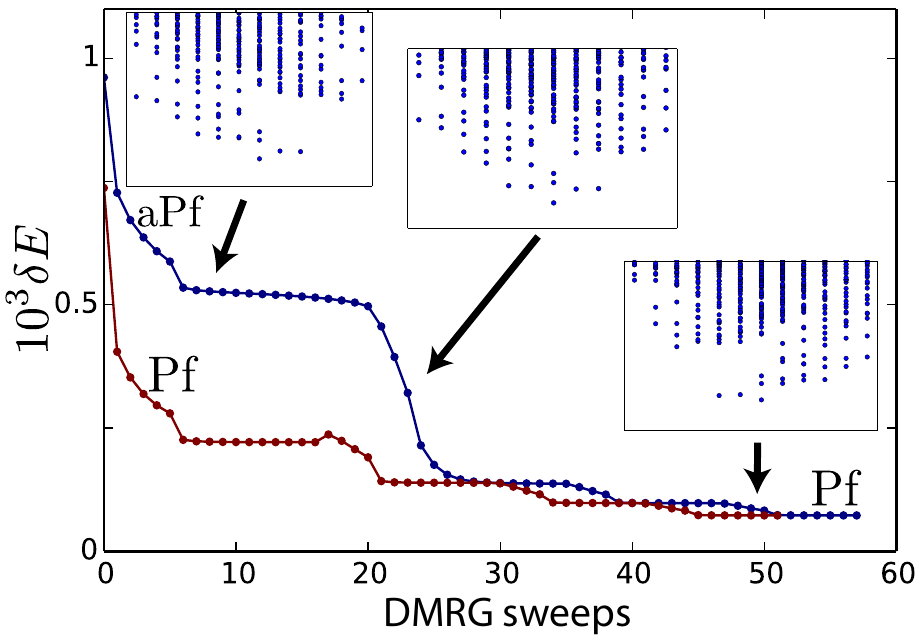} }
	\qquad
	\subfigure[\quad 3 Landau levels]{ \includegraphics[width=7.2cm]{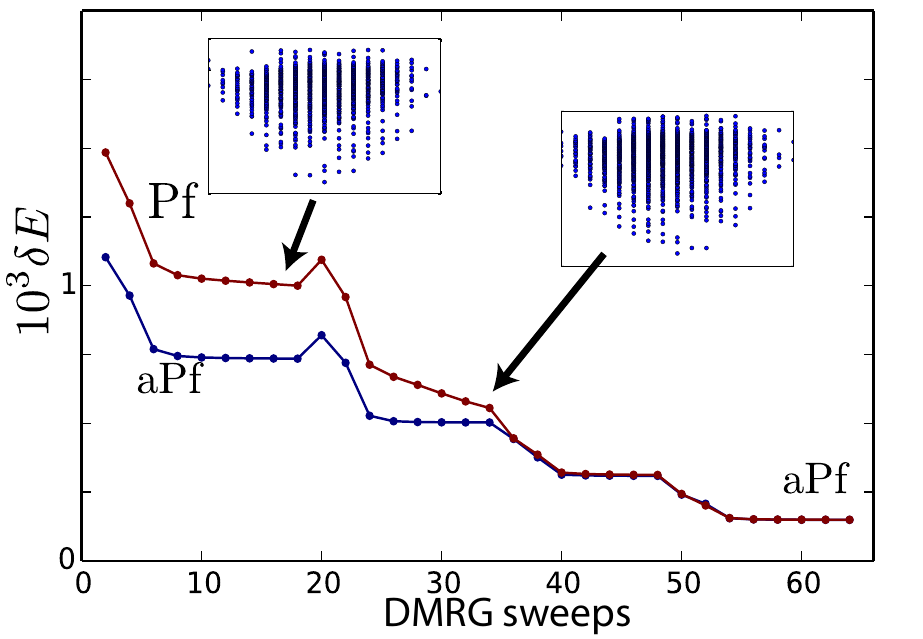} }
	\caption{%
	\textbf{Landau level mixing at} $\nu = 5/2$, $\kappa = 1.38$, $L = 20 \ell_B$. We find that when two LLs ($n = 0, 1$) are kept, as shown in panel {(a)}, the ground state is the Pfaffian state, while when three LLs ($n = 0, 1, 2$) are kept, as shown in panel {(b)}, the ground state is the anti-Pfaffian state. To verify this, in two separate runs we initialize the iDMRG simulation with model wavefunctions for the Pfaffian (red) and anti-Pfaffian (blue) states, for which exact matrix product state representations are known.
	The wavefunctions are then optimized using DMRG, and the energy is computed as the DMRG proceeds.
	\textbf{(a)} \textbf{For two LLs},  the MPS bond dimension is restricted to $\chi = 2600$ during the initial sweeps and is then increased in three steps up to its final value of 6300. The anti-Pfaffian appears to be metastable, with an energy per flux of the $\delta E \sim 0.5 \times 10^{-3}$ higher than the Pfaffian.
	The entanglement spectrum, shown in the inset, initially has a \emph{left} moving chirality, a clear signature of the anti-Pfaffian in our convention.
	As $\chi$ is allowed to increase (leading to the step-like energy decreases) the entanglement spectrum becomes a-chiral, then collapses to the Pfaffian. 
	The wavefunctions for the two runs are identical during the final sweeps of the DMRG.
	\textbf{(b)} \textbf{For three LLs}, the analysis is equivalent, but we find the anti-Pfaffian is preferred.}
\label{fig:PfaPf}
\end{figure*}

	The issue is clearly delicate, and may depend on the  sample details such as density, finite well width, subband mixing, disorder, etc. 
To be sure these have an important role in all types of experiments and, in particular, could account for the differences in the spin-polarization. 
Here we study only the benchmark case of a pure system with the Coulomb interaction and zero well width. 

There are currently two approaches to the problem of LL mixing.	
In the first `perturbative' approach, one integrates out the adjacent LLs to lowest order in $\kappa$ in order to derive an effective Hamiltonian for the $n=1$ LL, resulting in renormalized 2-body interactions and new 3-body terms.\cite{BisharaNayak2009, SimonRezayi2013,PetersonNayak,SodemannMacDonald}
This approach captures the effect of all LLs, and the resulting effective Hamiltonian can be studied using exact diagonalization of a single LL; once projected, $N_\Phi \sim 32\mbox{--}37$ fluxes can be studied.\cite{PetersonNayak,Nayak:3body:2014}
However, the expansion parameter $\kappa$ is of order 1, so it is not certain how accurate the perturbation theory is at the relevant values.  It is also a low density approximation in that the initial state 
of half-filled electrons is replaced by a very few electrons.
The magnitude of the generated terms seems to suggest the lowest order result is reasonable, but 
a more complex calculation to order $\kappa^2$ would be required to verify this.
When studying the effective Hamiltonian using exact diagonalization, there are finite size effects which are larger than those studied in this work.
Unfortunately, it is very burdensome to implement the effective Hamiltonian within infinite-DMRG, as the MPO required for a 3-body interaction is very large, so we have not yet pursued this approach. 

In the second approach, one studies the bare 2-body interaction in a truncated Hilbert space\cite{WojsQuinn2006} with higher LLs.
In the work of Ref.~\onlinecite{RezayiSimon-LLMixing}, for example, the bare 2-body interaction was studied in a Hilbert space which allowed for a limited number (say 1--3) of holes/electrons  in the $n = 0/2$ LLs. 
The approach we take in this work is similar, although we keep the entire Hilbert space of some finite number of LLs (up through $n = 5$).
This approach is entirely non-perturbative in $\kappa$, but neglects the effect of higher Landau levels.
Using the multicomponent iDMRG approach, we can keep the $n = 0, 1, 2, 3, 4$ LLs on cylinders up to circumference $L = 20 \ell_B$, which mitigates much of the finite size effects.

Clearly the perturbative and truncated Hilbert space approaches are complementary, as they make distinct approximations which are difficult to evaluate when using one method alone. 
We have made a preliminary investigation within the truncated Hilbert space approach, but for now must leave open the possibility that the truncation is unjustified.
All computations are performed at $L = 20 \ell_B$ and $\xiV = 5\ell_B$.

First we find the ground state in the full Hilbert space of
(I) $\NLL = 2$ with $n=0, 1$ LLs, and (II) $\NLL=3$ with $n=0, 1, 2$ LLs.
We fix $\kappa = 1.38$ for all the data presented here; a typical experimental value\cite{Xia12/5} that was studied numerically in Ref.~\onlinecite{RezayiSimon-LLMixing}.
We find definitive evidence that for (I), the Pf is preferred over the aPf, while for (II), the aPf is preferred, in agreement with Ref.~\onlinecite{RezayiSimon-LLMixing}.
The circumference $L = 20 \ell_B$ used here is nearly twice that of Ref.~\onlinecite{RezayiSimon-LLMixing}, which indicates finite size effects are not an issue.
Because it is believed that in the absence of LL mixing the system is poised at a first order transition between the Pf and aPf states,\cite{Peterson2008} the iDMRG may be susceptible to getting stuck in the wrong metastable state.
To address this metastability issue, in both cases (I) and (II) we run the iDMRG twice,
	first initializing the iDMRG with the exact MPS for the Pf state,\cite{ZaletelMong} and second with the exact MPS for the aPf state. 
The DMRG then proceeds to variationally optimize these two possibilities, as shown in Fig.~\ref{fig:PfaPf}, and we find the DMRG definitely chooses one or the other: if the run is initialized with the wrong ansatz, after several DMRG sweeps it eventually `tunnels' into the lower energy state.
This demonstrates there is no metastability issue and the iDMRG is reliable.

\begin{figure*}[t]
\includegraphics[width=2.\columnwidth]{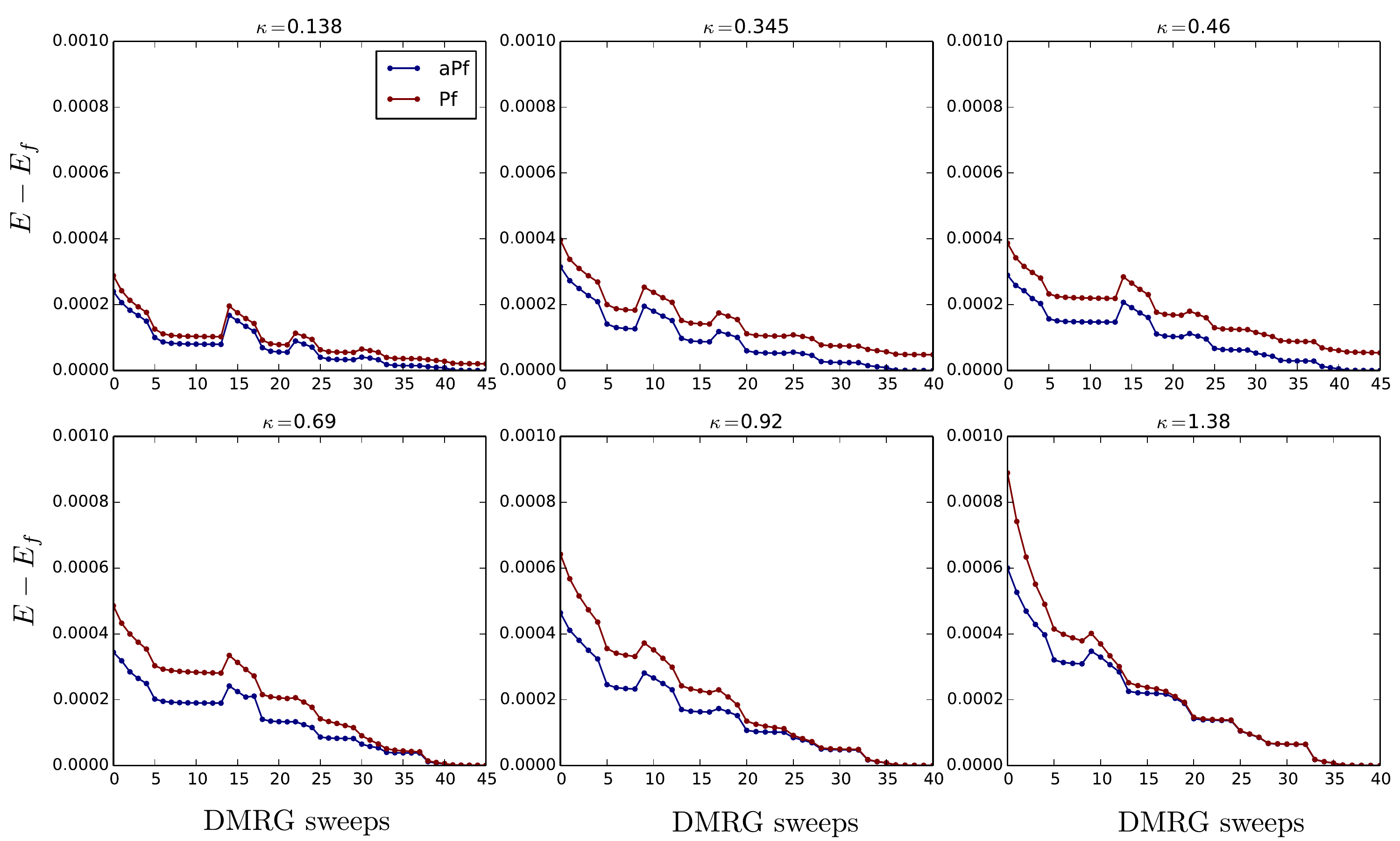} 
	\caption{The convergence of the energy per flux during the DMRG simulation when initialized from the anti-Pfaffian (blue) and Pfaffian (red) states. 
	As in Fig.~\ref{fig:PfaPf}(b), we simulate $L_x = 20 \ell_B$ while keeping the lowest $\NLL = 3$ Landau levels. However, we now consider a range of  Landau level mixing $\kappa = 1.38, 1.38 / 1.5, \dots $. For larger $\kappa$, the Pfaffian state tunnels into the anti-Pfaffian state. For smaller $\kappa$ the Pfaffian state remains metastable since the finite DMRG bond dimension discourages tunneling once the energetic splitting is too low. 
Regardless, there is an energetic splitting favoring the anti-Pfaffian state.
	}
\label{fig:Escan}
\end{figure*}

\begin{figure}
\begin{centering}
	\includegraphics[width=80mm]{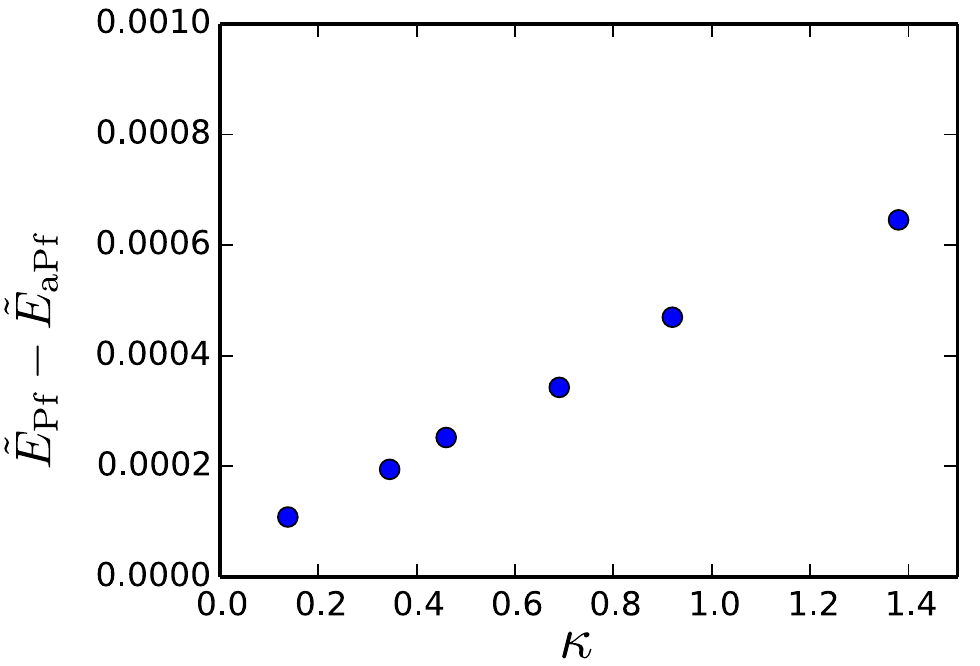}
	\caption{An estimate of the splitting per flux between the Pfaffian and anti-Pfaffian states as a function of $\kappa$.
To estimate the splitting, we calculate the energy difference per flux  $\tilde{E}_{\textrm{Pf}} - \tilde{E}_{\textrm{aPf}}$ between simulations initialized with the model Pfaffian and anti-Pfaffian states at the second sweep of the DMRG calculations, as shown in Fig.~\ref{fig:Escan}.
The energy splitting is metastable during the initial sweeps of DMRG; while it is not the true energetic splitting, it appears to be close to the true splitting found in exact diagonalization of a smaller torus.
As expected, we find a roughly linear dependence on $\kappa$.
}
	\label{fig:PfvaPf_splitting}
\end{centering}
\end{figure}

Second, we performed the same analysis for case (II), keeping the Landau levels $n = \{ 0, 1, 2\} $, but with decreasing values of  $\kappa = \{1.38, 1.38 / 2, \dots, 1.38 / 10 \} $.
We find that the aPf is preferred for all values of $\kappa$, as shown in Fig.~\ref{fig:Escan}.
Furthermore, the energy splitting between the Pf and aPf measured from the initial couple sweeps of the iDMRG scales as $\kappa$, at least qualitatively, as shown in Fig.~\ref{fig:PfvaPf_splitting}.
It is hard to assign a precise quantitative meaning to the energy splitting since  the Pf eventually tunnels into the aPf.
However, we believe the splitting $\tilde{E}_{\textrm{Pf}} -  \tilde{E}_{\textrm{aPf}}$ is qualitatively like the energy per flux separating two phases near a first order transition.
We find this strongly indicative that the aPf state remains preferred all the way to the $\kappa \to 0$ perturbative regime.  
Our finite-size studies on the torus corroborate this result. 
A good test case is when the ground state of the Coulomb potential for the half filling of the 1\textsuperscript{st}LL, without LL mixing, is doubly 
degenerate. This is not realized for even $N_e$, but it occurs for some odd sizes on a hexagonal
torus.\cite{PapicHaldaneRezayi}
For a 9-electron 1LL system we have confirmed in the same 3-LL model that even a very small $\kappa$ favors the aPf.

\begin{figure}
\begin{centering}
	\includegraphics[width=80mm]{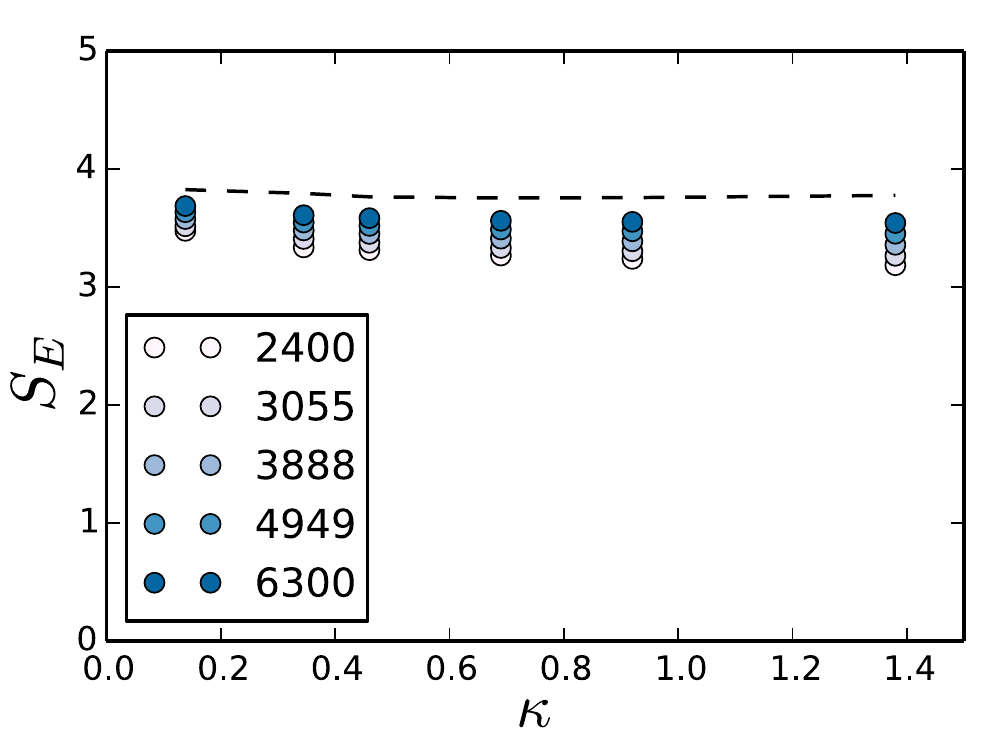}
	\caption{%
	The  entanglement entropy $S_E(\kappa)$ as a function of the Landau level mixing strength $\kappa$. We measure the entanglement entropy as the DMRG bond dimension is increased from $\chi = 2400$ to $6300$. The state is  not fully converged at $\chi = 6300$, so we extrapolate  $S_E$ in $1 / \chi$ to obtain an estimate of the converged result, shown as a dashed line. 
The resulting $S_E$ depends only weakly on $\kappa$, supporting a continuous dependence on $\kappa$ up to $\kappa = 1.38$.	
	}
	\label{fig:S_v_kappa}
\end{centering}
\end{figure}

The ground state appears to evolve smoothly (and weakly) with $\kappa$.
The entanglement entropy $S_E(\kappa)$ also depends very weakly on $\kappa$, as shown in Fig.~\ref{fig:S_v_kappa}, indicating an absence of a continuous phase transition between $\kappa = 0$ and $1.38$. 
The ``entanglement gap'' (the splitting between the lowest Schmidt weight corresponding to the edge conformal field theory and the lowest non-universal Schmidt weight present in the Coulomb state) also shows almost negligible decrease with $\kappa$.


\begin{figure}
	\includegraphics[width=80mm]{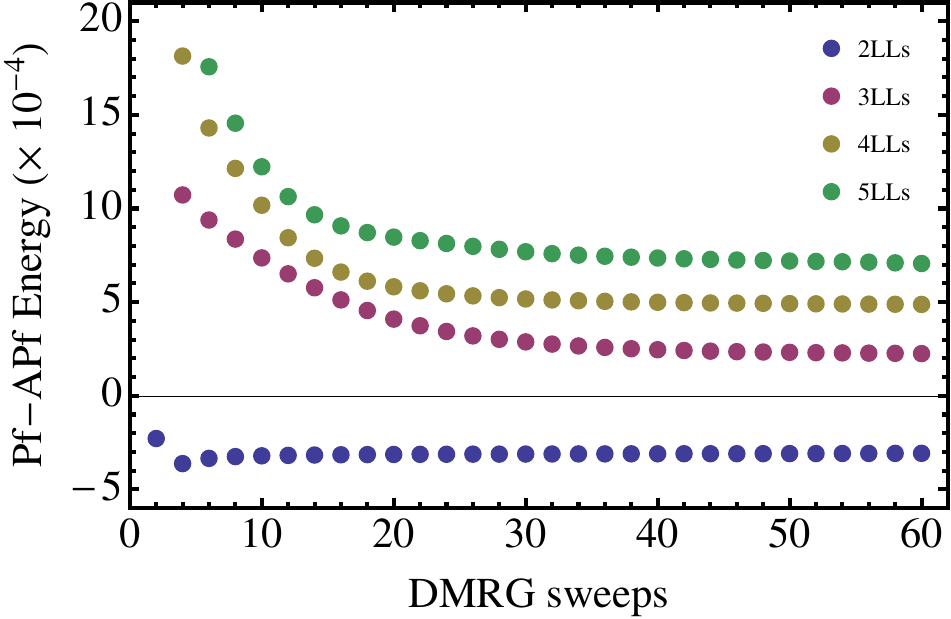}
	\caption{The splitting of the variational energies $E_{\textrm{Pf}} -  E_{\textrm{aPf}}$ per flux when the iDMRG is restricted to $\chi = 2000$.
		Again, the iDMRG is initialized with the Pf and aPf states, but for small $\chi$ the DMRG cannot tunnel between the two.
		The resulting energy splitting is measured while keeping the lowest $\NLL = 2, 3, 4, 5$ LLs.
		In agreement with the study of (I) and (II), only the $\NLL = 2$ case prefers the Pf.
	}
	\label{fig:moreLLs}
\end{figure}

Finally, we attempt to assess the accuracy of the truncated Hilbert space approach by including higher Landau levels up to $n = 4$.
Unfortunately, the resources required to converge the DMRG to the same level of precision as in cases (I) and (II) quickly become prohibitive. 
Instead, we restrict the iDMRG to a maximum of $\chi = 2000$ Schmidt states. 
Again initializing the DMRG with both the Pf and aPf states, we find the DMRG does \emph{not} tunnel between the Pf and aPf, because the small $\chi$ generates a barrier which prevents the tunneling.
Since the situation is metastable, we can measure two variational energies $E_{\textrm{Pf}}$ and $E_{\textrm{aPf}}$. 
In Fig.~\ref{fig:moreLLs}, we plot the splitting $E_{\textrm{Pf}} -  E_{\textrm{aPf}}$ when 2, 3, 4 and 5 LLs are kept. 
For $\NLL = 2$ LLs, the Pf is preferred as before, while for $\NLL > 2$ the aPf.
The aPf becomes more strongly preferred when keeping higher $\NLL$. 
(We note that the energy difference shown in Fig.~\ref{fig:moreLLs} is sensitive to $\chi$, and thus the plots should be understood as being only qualitative.)
However, while unlikely, we cannot rule out the possibility of the sign eventually switching again.

\subsection{Discussion}
	The results here should be combined with the perturbative approach to reach a trustworthy conclusion.
We note that Pakrouski \textit{et al.}\ have pursued the pertubative approach but concluded that the Pfaffian is preferred.\cite{Nayak:3body:2014}
In light of this discrepancy, more work must be done to resolve the state at $\nu = 5/2$.

A first test is to carefully check for agreement between the truncated Hamiltonian approach and the perturbative approach in the $\kappa \to 0$ limit; if they disagree, then presumably  truncating  higher LLs is unjustified and should not be pursued further. 
While our results appear to disagree with those of Ref.~\onlinecite{Nayak:3body:2014} in this regime, the effective Hamiltonian used in their torus calculation was calculated using perturbation theory on an infinite plane, which introduces some uncertainties.
Furthermore, the splitting between the Pfaffian and anti-Pfaffian states on the torus does not scale extensively with system size, indicating there may still be large finite size effects.

A second test would be to restrict the perturbative approach to include only the lowest $\NLL$ LLs and exactly diagonalize the resulting effective Hamiltonian.
If the results depend strongly on $\NLL$ (for example, preferring the Pfaffian for $\NLL = 2$, the aPf for $\NLL = 3, 4, 5$, but the Pf again for $\NLL = \infty$), then the truncation approach would appear to be unjustified.
If the truncation approach passes both these tests, then the results shown here provide strong evidence that the aPf is preferred up to $\kappa \sim 1$ and at large system sizes $L \sim 20 \ell_B$, and further investigation into how the finite well width could be used to stabilize the phase would be worthwhile.
If the truncation proves to be unjustified, but finite size effects limit the reliability of the effective Hamiltonian ED, then one could tediously construct the MPO for the effective 3-body terms for use in iDMRG.

\section*{Acknowledgment}
We acknowledge J.~E.~Moore, C.~Nayak, S.~Simon, and S.~Parameswaran, for helpful comments and conversations.
We are indebted to Z.~Papi\'{c} for advice and triple checks to our data.
MPZ is grateful for the support of NSF Grant No.~DMR-1206515;
RM acknowledges funding from the Sherman Fairchild Foundation;
EHR is supported by DOE Grant No.~DE-SC0002140.
MPZ and RM gratefully acknowledge support from the visitors program of the Max Planck Institute for the Physics of Complex Systems, Dresden.

\bibliography{qbench}

\appendix

\section{Example of the MPO compression}
\label{app:mpo_example}
Here we give a concrete example of how the matrices $(A, B, C, D)$ described in Sec.~\ref{sec:MPOcompress} are used to construct an MPO.
For simplicity we show only the $m = 0$ sector with a single component, with interaction of the form $\sum_{k \geq0} V_k \hat{n}_{r} \hat{n}_{r+k}$.
The MPO is encoded in a matrix of operators $\hat{W}$, which for this case takes the block form
\begin{align}
	\hat{W} = \left(\!\begin{array}{c|c|c} \mathds{1} & C\hat{n} & D\hat{n}^2 \\
	\hline 
	0 &  A & B\hat{n} \\
	\hline 
	0 & 0 & \mathds{1} \end{array}\!\right) .
\end{align}
(Here $\mathds{1}$ is the identity operator.)
In this simplified example, $A$ is a $\Lambda \times \Lambda$ matrix, $B$/$C$ are column/row vectors of length $\Lambda$, and $D$ is a scalar, satisfying Eq.~\eqref{eq:Vdecomp}.
Thus the MPO has size $(\Lambda + 2) \times (\Lambda + 2)$.

As an example, an interaction of the form $V_k = \cos(\beta k) e^{- \alpha k}$ may be written as a sum of two exponentials $V_k = \tfrac{1}{2} ( e^{ -z k} + e^{ - \bar{z} k})$, where $z = \alpha + i \beta$.
Notice that the MPO has a `gauge' redundancy of the form $(A, B, C, D) \to (G^{-1} A G, G^{-1} B, C G, D)$ for an invertible matrix $G$.
This can be used to bring $A$ to various canonical forms.
Choosing $A$ to be diagonal, we have
\begin{align}
	\hat{W} = \left(\!\begin{array}{c|cc|c} \mathds{1} & e^{-z}\hat{n} & e^{-\bar{z}}\hat{n} & \hat{n}^2 \\
	\hline 
	0 & e^{ - z} & 0 & \hat{n}/2 \\0 & 0 & e^{ - \bar{z}} & \hat{n}/2 \\
	\hline 
	0 & 0 & 0 & \mathds{1} \end{array}\!\right) .
\end{align}
However, this is numerically sub-optimal since the entries are complex, despite $V_k$ being real.
Instead we can use the real block-Schur form.
Defining  constants
\begin{align}
c =  \cos(\beta) e^{- \alpha}, \quad \quad s =  \sin(\beta) e^{- \alpha},
\end{align}
the gauge freedom $G$ allow us to write
\begin{align}
	\hat{W} = \left(\!\begin{array}{c|cc|c}\mathds{1} & c  \hat{n}  & -s  \hat{n} & \hat{n}^2 \\
	\hline 
	0 & c & -s  & \hat{n} \\0 & s & c & 0 \\
	\hline 
	0 & 0 & 0 & \mathds{1}\end{array}\!\right).
\end{align}

Generically,  if the Hamiltonian is real in the chosen single-particle basis, the $A$ produced by the block-Hankel compression can always be brought to the real block-Schur form with $2\times2$ blocks along the diagonal.
This should be the case for quantum Hall systems with a $180^\circ$ rotational symmetry and time-reversal invariant interactions.

\end{document}